# The Effect of Real Estate Auction Events on Mortality Rate

Cheoljoon Jeong

Credit Bureau Business Division
National Information and Credit Evaluation, Seoul, South Korea 07237
cjjeong@niceinfo.co.kr

**Abstract** This study has investigated the mortality rate of parties at real estate auctions compared to that of the overall population in South Korea by using various variables, including age, real estate usage, cumulative number of real estate auction events, disposal of real estate, and appraisal price. In each case, there has been a significant difference between mortality rate of parties at real estate auctions and that of the overall population, which provides a new insight regarding utilization of the information on real estate auctions. Despite the need for further detailed analysis on the correlation between real estate auction events and death, because the result from this study is still meaningful, the result is summarized for informational purposes.

**History** published on April 15, 2017

## I. Introduction

This analysis began from the controversial suicide issues in relation to insurance coverage throughout 2016. If a specific relationship between the characteristics of suicide deaths and retained information is revealed, it could contribute to further analysis that, even if indirectly, determines the characteristics of suicides. Accordingly, this study first analyzed the information on real estate auctions from retained information. However, before exploring aforementioned relationship, this study first compared the general mortality rate of the parties in real estate auctions with that of the overall population in South Korea.

**Controversies over Insurance Coverage on Suicide in Life Insurance Industry**

Controversies over insurance coverage on suicide have persisted in the life insurance industry for more than a decade. In the early 2000s, some insurance companies began to sell insurance products containing a disaster death benefit rider, which allows claims for those who have committed suicide two years after the subscription to that insurance. The issue in this product lay in the disaster death benefit rider prescribing that the death benefit due to disasters should be twice or three times that for general death. The insurance company in this case delayed the payment of the death benefit, arguing that "suicide may not be considered as a disaster." The Supreme Court in 2016 held that "The death benefit may not be paid after the statute of limitations has expired," and accordingly, the insurance company did not pay the death benefit. However, the Financial Supervisory Service maintained its position that an insurance company should pay the death benefit, even if based on inappropriate terms and conditions, because the benefit was promised with the client, resulting in the heated debate over insurance coverage on suicide. This issue has inevitably attracted continuous attention from the life insurance industry, including debate over what the characteristics of suicide death are.

**New Approach to Real Estate Auction Information**

Typically, financial institutions or credit bureaus may use information on real estate auctions to request debt repayments through the court before the end date of the debt repayment request[1] if a debtor is an auctioneer, or to subject the claim right against the debt repayment to the (provisional) seizure to easily seek payment of non-performing loans (NPLs) with respect to a corresponding debtor if a debtor was the owner of auctioned items. Furthermore, the information on real estate auctions is also used to re-classify asset quality regarding the credit below the 'substandard' level[2] as set by financial institutions. According to a recently emerging view on the information on real estate auctions, in addition to utilizing the information for the purposes of debt collection and reclassification of asset quality, banks and credit card companies may be able to utilize the information on auctions to determine the risk of borrowers in reviewing a loan application, and establish the strategy based on that risk. For example, in the case of Card Company A[3], insolvency in a new enrollment to a credit card service was approximately double among those who were in the middle of the auction process, or who had the auction process less than a year before the enrollment, than those who were not party to any auction. This result facilitated a new approach to the information on real estate auctions. Together with this background information, the significant difference in mortality rate between parties at real estate auctions and the overall population in the country, which has been identified through this investigation, leads to a more interesting view. Despite the need for closer analysis on the correlation between real estate auctions and death, this analysis opens new opportunities for utilization of information on real estate auction. This study first examines the general trends of the real estate auction market and introduces the analytical process and the results. Subsequently, this study describes the difference in the mortality rate between parties at real estate auctions and the overall population in the country in various viewpoints, including age, real estate usage, cumulative number of real estate auction events, disposal of real estate, and appraisal price.

## II. Trends in Real Estate Auctions

**Distribution of Real Estate Auction Cases by Year**

The annual average number of real estate auction cases is 79,746. In 2012–2014, more real estate auction cases occurred than in any other years, though the number of such cases has been decreasing since 2013, and the number is returning to the average level in terms of the overall average.

---

[1] **End date of debt repayment request**: A debt repayment request is a method in which a creditor other than the distraining creditor participates in execution and is reimbursed in compulsory execution. In accordance with the Civil Act, the Commercial Act, and other laws, a creditor having a claim for preferential payment, a creditor having a certified copy of execution clause, or a creditor who has received a provisional seizure order after the registration recording the auction commencement decision, may petition the court for a debt repayment. The end date of debt repayment request (i.e., the final day available for a creditor to request a debt repayment) is determined as one day after two to three months from the date of the auction commencement decision by the court prior to the debt repayment request period.

[2] **Credit below the substandard**: Financial institutions regularly perform asset quality classification if presented with doubtful accounts. Financial institutions categorize credit into five levels: normal, special mention, substandard, doubtful, and estimated loss, based on the current status of the credit. The rating below the substandard refers to substandard, doubtful, and estimated loss, which are collectively known as non-performing loans (NPLs). In other words, NPL is a term combining 'estimated loss' (that requires loss treatment due to unavoidable insolvency), 'doubtful' (in which the loss is expected but not certain), and 'substandard' (i.e., recoverable through sales of collateral).

[3] **Validity analysis of auction information on new enrollments in credit card services (September 8, 2016)**: Non-performing target was defined as a customer with a record of default at the Korea Credit Information Services or at the credit bureau, or for whom the maximum days overdue is 60 or higher, within 12 months from enrollment date.

**Composition of Uses of Real Estate for Auction**

The top 10 types among the 64 types of real estate subject to auction include apartments, multi-family houses (villas), patty, field, forestry, residential houses, commercial buildings, land, factory, and residential officetels. In particular, apartments and villas account for 31.14%.

**Sale Value Ratio Trend by Year**

The sale value ratio, which is the ratio of the sale price to the appraisal value (appraisal price) for a sold lot shows variation over time. A higher sale value ratio indicates that a lot at the auction was sold at a higher price than at other times. In addition, the trend of increasing sale value ratio since 2012 suggests that values of lots at auction, as evaluated by the market, have been increasing.

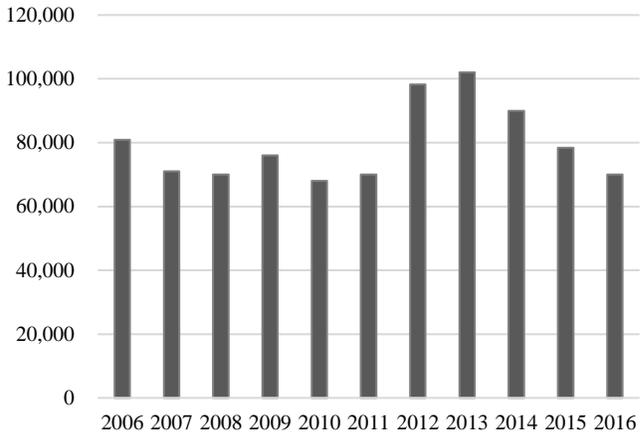

Figure 1: Distribution of Real Estate Auction Cases by Year

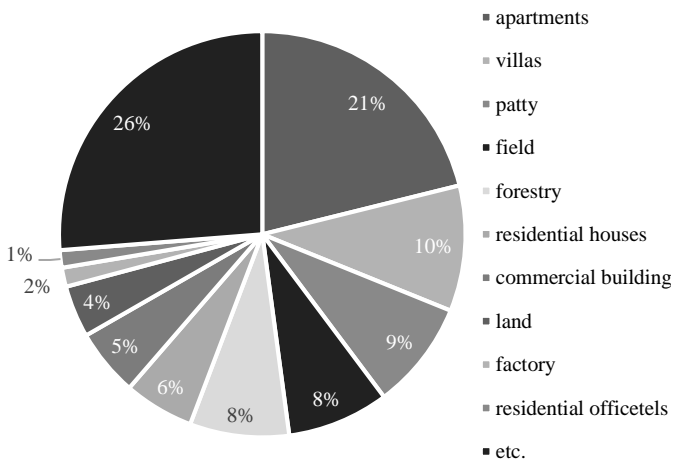

Figure 2: Composition of Uses of Real Estate for Auction

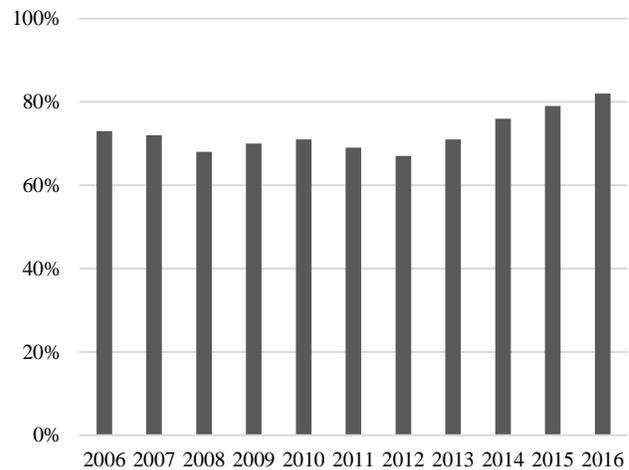

Figure 3: Sale Value Ratio Trend by Year

**Number of Parties[4] by Auction Lot**

The parties in a real estate auction can be divided into lot owners and auctioneers. Lot owners include tenants, subrogating people, surrogate applicants, lienholders, jeonse holders, people with the right of superficies, people with provisional disposition, people with preliminary registration, people with provisional seizure, people with seizure, right holders with compulsory auction, right holders with voluntary auction, and creditors; and auctioneers include single owners, owners under joint tenancy, and owners under tenancy in common. Since 2006, the number of auction lots has amounted to 1,044,055, the number of auctioneers per auction lot was 4.7 people, and the number of owners per auction lot was 2.6 people.

---

[4] **Number of Parties**: The number of people/corporations in one Certificate of Real Estate Register (Registered Copy) after deleting duplicate names thereof.

| Type of Parties | Number of Parties | Number of Parties per Auction Lot |
|---|---|---|
| Auctioneers | 4,940,381 | 4.7 |
| Lot Owners | 2,662,395 | 2.6 |
| Total | 7,602,776 | 7.3 |

Table 1: Number of Lot owners and Auctioneers per Auction Lot

## III. Comparison of Mortality between Parties at Real Estate Auctions and Overall Population in South Korea

### 1. Requirements for Analysis

**Category of Analysis Targets**

As of the end of 2016, the information on the auctioneers including single owners, owners under joint tenancy, and owners under tenancy in common, was extracted from the information on real estate auctions. In particular, because a share auction may be initiated without using an actual auction, any subjects vulnerable to direct damage due to auctions should be distinguished. For this reason, debtors were selected as analysis targets among owners under tenancy in common from the court auction information website (http://www.courtauction.go.kr). Ultimately, 513,166 people were selected as the analysis target, and the data were sorted solely based on birth date and name. (Because one of the sources of real estate auction information is the Certificate of Real Estate Register, the data was sorted based on the first six digits of the resident registration number as well as the name indicated on the certificate).

| Type of Target | Frequency | Proportion | Description |
|---|---|---|---|
| Target of Interest | 9,834 | 1.9% | An auctioneer who has experienced real estate auction within three years prior to the date of death |
| Targets of Non-interest | 503,332 | 98.1% | An auctioneer who has not experienced real estate auction or was still alive within the three years prior to the date of death |
| Analysis Targets (Auctioneers) | 513,166 | 100.0% | Auction debtors among single owners, owners under joint tenancy, and owners under tenancy in common based on birth date and name |

Table 2: Category of Analysis Targets

**9,834 Targets (1.9%) of Interest among 513,166 Analysis Targets**

The analysis targets (513,166 people) were matched with the death information from the Ministry of the Interior[5] to identify the dead among the analysis targets based on "birth date + name." The number of targets of interest was 9,834 people from the 513,166 total, comprising approximately 1.9%; the remaining 503,332 people (98.1%), were the auctioneers who have not experienced real estate auctions within three years prior to the date of death, or who were still alive. This study has compared and analyzed the mortality rate between the targets of interest (9,834 people) and the overall population in South Korea.

---

[5] **Death information from the Ministry of the Interior**: If the resident registration number (similar to the social security number in U.S.) that is retained by a credit bureau is submitted to the Korea Credit Information Services, the Korea Credit Information Services returns the date of death for those identified by the Ministry of the Interior.

**Definition of Terms: Mortality Rate of interest, Average Mortality Rate, and Potential Risk**

The "mortality rate of interest" is the number of auctioneers who have experienced real estate auctions within three years prior to the date of death divided by the total number of auctioneers, and the "average mortality rate," based on the Population Projections for Korea by Statistics Korea (KOSTAT), is the annual mortality rate with respect to the corresponding population by age (10-year interval) and region (city/province level), multiplied by 3, which represents the mortality rate for 3 years. The "potential risk" is the value of mortality rate of interest divided by the average mortality rate, which indicates how much higher the mortality rate of the parties is (which is expected to be associated with real estate auctions) than the general mortality rate.

| Terminology | Description |
| --- | --- |
| Mortality Rate of Interest | The number of auctioneers who have experienced real estate auction within three years prior to the date of death divided by the total number of auctioneers |
| Average Mortality Rate | An estimated value of mortality by age and region (city/province level) based on the data of Statistics Korea |
| Potential Risk | The value of mortality rate of interest divided by the average mortality rate (= mortality rate of interest /average mortality rate), which indicates the relative magnitude of potential risk of death |

Table 3: Definition of Terms: Mortality Rate of interest, Average Mortality Rate, and Potential Risk

## 2. Analysis Results

**In the lower age range, the potential risk of targets of interest is relatively higher than that of targets of non-interest**

In the age range of 30s and 40s, the ratio of mortality rate of interest to average mortality rate (potential risk) is approximately 1.9 times higher than that of the general case, and 1.1 times higher than the average of the overall population. In the age range of 50s and 60s, the potential risk was determined as 1.4 times and 1.3 times for respective cases, both of which are higher than the overall average of 1.1 times. In the age range of 10s and 20s, because the number of deaths is low to calculate reliable statistical significance, the data on this age range should be excluded from the interpretation. In this case, it can be noted that the potential risk of the targets of interest in the lower age range tends to be higher than that in the general case.

**Potential risk is higher in multi-family houses (villas) among types of housing; land, paddy, and field among types of real estate uses; and factory in commercial facilities**

In terms of the types of real estate uses, the potential risk of residential houses over land; residential houses and land over commercial facilities; residential houses over land; and land over residential houses was found to be higher than the average for the corresponding age ranges of 30s, 40s, 50s, and 60s, and higher than the average of the entire auctioneer group, of 1.1 times. Overall, cases where the real estate subject to auction is a residential house, regardless of age, showed a higher potential risk than the case of non-residential properties. However, the case where the type of real estate uses is land in the age range of 60s, exhibited a higher potential risk than the case of residential houses.

| Age | Analysis Targets | Number of Deaths | Ratio of Mortality Rate of Interest | Average Mortality Rate | Date Interval between Auction and Date of Death | Potential Risk (Times) |
|---|---|---|---|---|---|---|
| 10s | 3,140 | 3 | 0.10% | 0.50% | 704 | 2.0 |
| 20s | 14,574 | 33 | 0.23% | 1.30% | 581 | 1.8 |
| **30s** | **66,173** | **281** | **0.42%** | **2.20%** | **527** | **1.9** |
| **40s** | **162,725** | **1,562** | **0.96%** | **5.00%** | **525** | **1.9** |
| 50s | 157,862 | 2,413 | 1.53% | 1.09% | 521 | 1.4 |
| 60s | 69,324 | 2,110 | 3.04% | 2.33% | 526 | 1.3 |
| 70s | 30,721 | 2,251 | 7.33% | 7.08% | 551 | 1.0 |
| 80s | 6,810 | 1,055 | 15.49% | 23.81% | 527 | 0.7 |
| 90s | 1,837 | 126 | 6.86% | 59.42% | 473 | 0.1 |
| Total | 513,166 | 9,834 | 1.92% | 1.79% | 530 | 1.1 |

Table 4: Ratio of Mortality Rate of Interest and Potential Risk of Targets of Interest by Age

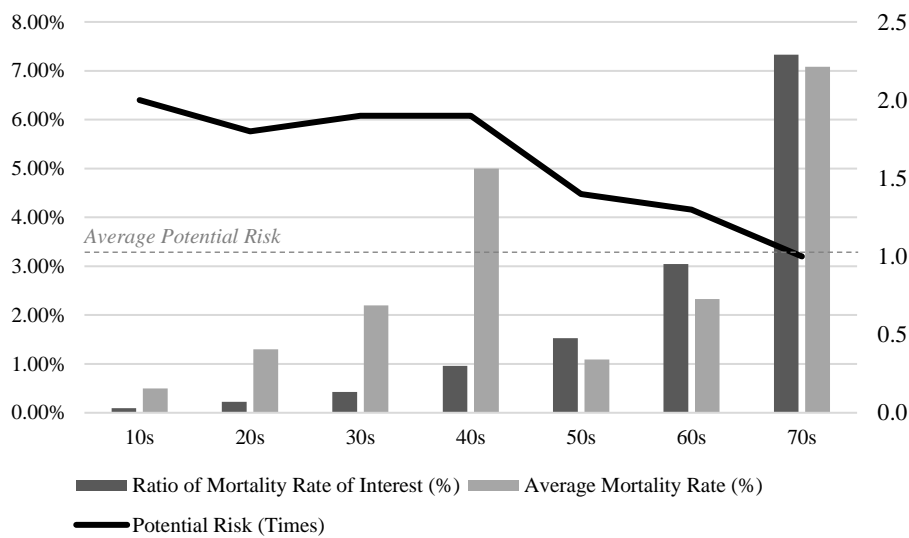

Figure 4: Ratio of Mortality Rate of Interest and Potential Risk of Targets of Interest by Age

The previously described real estate uses were determined by grouping various usages in detail. A closer look at the real estate use types shows that the potential risk was determined as higher for the upper age range having farmhouses among the housing types, and for lower age ranges possessing multi-family apartments, residential officetels, or villa type apartments. Despite the low mortality, in terms of land, the potential risk was determined as higher for the upper age range having agriculture-related lands including barns, hybrids, orchards, farm-related facilities, pasture lands, and ditches, and for the lower age range possessing land, paddy, field, or forestry. In the case of commercial facilities, although the mortality is lower than that for housing and land, the potential risk was higher than the average for those who have commercial facilities including commercial buildings, factories, gas stations, religious facilities, and factory sites. The age range of people owning these real estate types falls between 40s and 60s.

| Category | Types of Real Estate Uses | Age | | | | | |
|---|---|---|---|---|---|---|---|
| | | 20s | 30s | 40s | 50s | 60s | 70s |
| Potential Risk (Times) | Residential Houses | 1.7 | 2.0 | 2.0 | 1.6 | 1.4 | 1.2 |
| | Land | 2.7 | 1.8 | 2.0 | 1.5 | 1.5 | 1.1 |
| | Commercial Facilities | 0.9 | 1.2 | 1.5 | 1.2 | 1.3 | 1.0 |
| | Etc. | - | 1.8 | 1.4 | 1.0 | 0.8 | 0.8 |
| | No-File | - | - | - | - | - | - |
| Analysis Targets | Residential Houses | 8,050 | 40,346 | 92,599 | 76,700 | 30,296 | 13,214 |
| | Land | 4,106 | 12,211 | 34,005 | 40,273 | 20,106 | 9,636 |
| | Commercial Facilities | 835 | 4,157 | 12,542 | 13,419 | 5,449 | 1,907 |
| | Etc. | 1,572 | 9,441 | 23,532 | 27,403 | 13,443 | 5,953 |
| | No-File | 11 | 18 | 47 | 67 | 30 | 11 |
| Number of Deaths | Residential Houses | 17 | 178 | 927 | 1,265 | 992 | 1,070 |
| | Land | 15 | 55 | 377 | 693 | 708 | 738 |
| | Commercial Facilities | 1 | 11 | 95 | 167 | 163 | 128 |
| | Etc. | - | 37 | 163 | 288 | 247 | 315 |
| | No-File | - | - | - | - | - | - |
| Average Potential Risk | | 1.8 | 1.9 | 1.9 | 1.4 | 1.3 | 1.0 |
| Total Analysis Targets | | 14,574 | 66,173 | 162,725 | 157,862 | 69,324 | 30,721 |
| Total Number of Deaths | | 33 | 281 | 1,562 | 2,413 | 2,110 | 2,521 |

Table 5: Potential Risk by Age and Types of Real Estate Uses

**Mortality rate of interest and potential risk are low when auction experience is high**

The cumulative number of auction cases for the targets of interest is determined from the auction commencement date to the date of death, while that for the targets of non-interest is determined from the auction commencement date to the end of 2016. In this case, as the cumulative number of auction cases increases, the mortality rate of interest and potential risk tend to become lower. It seems intuitive that the mortality rate of interest and potential risk would be high if the auction experience is extensive; however, the analysis results support the opposite cases.

| Cumulative Number of Auction Cases | Analysis Targets | Number of Deaths | Ratio of Mortality Rate of Interest | Average Mortality Rate | Date Interval between Auction and Date of Death | Potential Risk (Times) |
|---|---|---|---|---|---|---|
| 1 | 397,003 | 8,117 | 2.04% | 1.83% | 523 | 1.1 |
| 2 | 80,177 | 1,269 | 1.58% | 1.69% | 564 | 0.9 |
| 3 | 21,992 | 302 | 1.37% | 1.62% | 577 | 0.8 |
| 4 | 7,369 | 92 | 1.25% | 1.53% | 526 | 0.8 |
| 5+ | 6,625 | 54 | 0.82% | 1.48% | 587 | 0.6 |
| Total | 513,166 | 9,834 | 1.92% | 1.79% | 530 | 1.1 |

Table 6: Ratio of Mortality Rate of Interest and Potential Risk of Targets of Interest by Cumulative Number of Auction Cases

**Mortality rate of interest and potential risk are high when lots have been sold**

Even if an auction is initiated, the real estate is not necessarily successfully sold. Before the auction is closed, the case could be withdrawn, dismissed, or rejected, and the case may be changed, or suspended. Furthermore, multiple cases could be annexed, the auction could be passed in, or resumed, and the lot could be resold. The actual completion of sale of the lot via auction can be determined if the notices are released at the court auction information website, including sale, balance submission (the completed sale of real estate because the debt repayment is distributed to the right holders and the balance was submitted prior to the end date of debt repayment request), and completed debt repayment. According to the analysis results, the mortality rate of interest was determined as higher by approximately 0.4% in the case when the real estate was sold than in the case when the real estate was not sold, which was higher than the overall average of potential risk with 1.1 times. In other words, the mortality rate of interest and the potential risk depend on whether the real estate is sold.

| Category | Analysis Targets | Number of Deaths | Ratio of Mortality Rate of Interest | Average Mortality Rate | Date Interval between Auction and Date of Death | Potential Risk (Times) |
|---|---|---|---|---|---|---|
| Sold | 329,329 | 6,787 | 2.06% | 1.74% | 540 | 1.2 |
| Not Sold | 183,837 | 3,047 | 1.66% | 1.90% | 508 | 0.9 |
| Total | 513,166 | 9,834 | 1.92% | 1.79% | 530 | 1.1 |

Table 7: Ratio of Mortality Rate of Interest and Potential Risk of Targets of Interest in Terms of Whether Real Estate is Sold or Not

**Mortality rate of interest and potential risk are high when appraisal price is low**

The mortality rate of interest and the potential risk were investigated by appraisal price[6], which is the value of the real estate subject to auction (owned by the target of interest). The range of the appraisal was divided into seven intervals from KRW 50 million or less to KRW 2 billion or more, and the part indicated as KRW 0 was excluded from the analysis. According to the results examining the mortality rate of interest and the potential risk by intervals, if the appraisal price is low, the mortality rate of interest as well as the potential risk tends to be relatively high. In particular, the mortality rate of interest for the intervals for KRW 100 million or less was by 0.48% higher than for the intervals for KRW 100 million or more; the potential risk was determined as 1.3 times higher in the interval ranging from KRW 50 million to KRW 100 million, and 1.2 times higher in the interval of KRW 50 million or less, than the overall average (1.1 times). When low-value real estate is subject to auction, the potential risk to the party is higher than the overall average.

In addition, the period from the auction commencement date to the date of death was 572 days when the appraisal price was KRW 2 billion or more, which exceeded the average of 530 days. If the appraisal price was KRW 20 billion or more, the mortality rate of interest is also lower than the average mortality rate, and accordingly, the potential risk is lower than the average of 1.1 times among all auctioneers. It is interesting to note that when the real estate valued as high is subject to auction, the mortality rate of interest for the parties is lower than the overall average.

| Appraisal Price (KRW) | Analysis Targets | Number of Deaths | Ratio of Mortality Rate of Interest | Average Mortality Rate | Date Interval between Auction and Date of Death | Potential Risk (Times) |
|---|---|---|---|---|---|---|
| 0 million | 82,269 | 1,192 | 1.45% | 2.02% | 430 | 0.7 |
| ~ 50 million | 124,407 | 3,265 | 2.62% | 2.21% | 552 | 1.2 |
| 50 ~ 100 million | 91,423 | 1,920 | 2.10% | 1.63% | 549 | 1.3 |
| 100 ~ 300 million | 133,457 | 2,159 | 1.62% | 1.49% | 533 | 1.1 |
| 300 ~ 600 million | 49,965 | 789 | 1.58% | 1.51% | 533 | 1.0 |
| 600 million ~ 1 billion | 16,784 | 267 | 1.59% | 1.62% | 530 | 1.0 |
| 1~2 billion | 9,837 | 170 | 1.73% | 1.76% | 530 | 1.0 |
| 2 billion ~ | 5,024 | 72 | 1.43% | 1.91% | 572 | 0.7 |
| Total | 513,166 | 9,834 | 1.92% | 1.79% | 530 | 1.1 |

Table 8: Ratio of Mortality Rate of Interest and Potential Risk of Targets of Interest by Appraisal Price

---

[6] **Appraisal price**: An appraisal price is a real estate value assessed by a professional appraiser, which is the basis for determining the minimum sale price upon the execution of an auction. Although an execution court is not required to set the value determined by the appraiser as the minimum sale price, the value determined by the appraiser is accepted as the minimum sale price in practice. The appraisal price assessed in this manner deviates from the actual market price or sale price of the real estate; however, the appraisal price was used as significant means to measure the price in this analysis because the appraisal price is determined according to trends in market price.

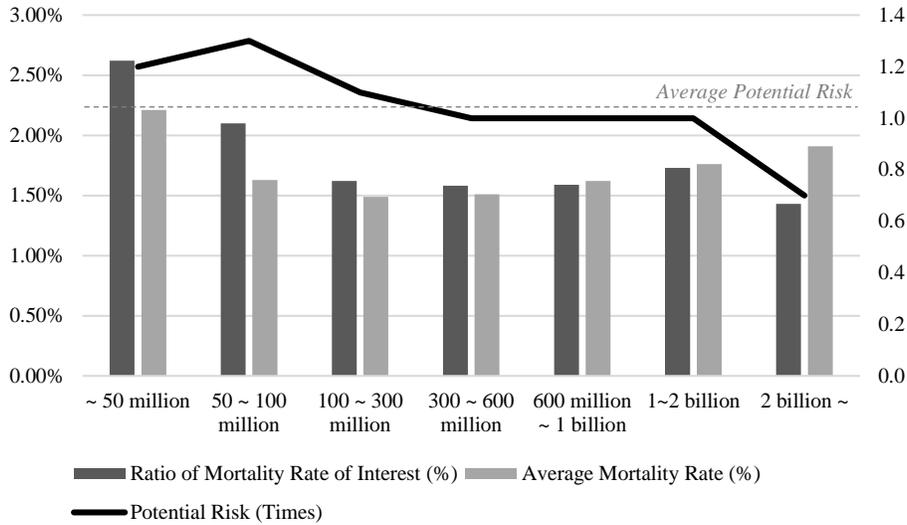

Figure 5: Ratio of Mortality Rate of Interest and Potential Risk of Targets of Interest by Appraisal Price

According to the appraisal price of the real estate by age, the potential risk is determined as high for the 30–80 year age range, with KRW 100 million or less as the appraisal price of the real estate In particular, in the interval ranging from KRW 500 million to KRW 1,000 million (having the higher potential risk), those who are in the lower age range showed an even higher potential risk.

## IV. Conclusion

This study has so far investigated the difference in the mortality rate between the targets of interest and the overall population in South Korea across different variables, including age, real estate usage, cumulative number of real estate auction events, disposal of real estate, and appraisal price. When the targets of interest are in the lower age range, the real estate in possession is closely related to everyday life, the number of cumulative auction cases is low, the target real estate was actually sold, and the appraisal price of the real estate is low, the potential risk interestingly was determined as higher than the overall average. It should be noted that this analysis does not guarantee any sequential or causal relationship between real estate auction and death. When we look at social phenomena, we often assume that if two phenomena move together in a forward or reverse direction, there will be an "interactive relationship" between the two phenomena (i.e., the well-known post hoc ergo propter hoc fallacy). A well-known example is that in the 1990s, the sales of diapers and beers showed similar trajectories at Wal-Mart, a U.S. large discount store. Without any follow-up analysis, no sequential or causal relationship between diapers and beers can be determined; it can only be concluded that the trends of the two variables evolved in a similar manner.). The results of this analysis also suggest that the possible association between the two variables could be "established as a fact" only with additional in-depth analysis as well as examining various information and phenomena together. Various types of information including public information have recently been disclosed through the Internet, and data collection technology has been evolving rapidly. This is referred to as the big data era. Thus, credit bureaus have continuously been monitoring the possibility of utilizing alternative data as an auxiliary indicator of conventional credit information. Among credit information, real estate information consists of various types of information, including registration application cases, market prices, a Certificate of Real Estate Register, and a Comprehensive Certificate of Real Estate, as well as auction information, requiring additional analysis regarding each type of information and how they interrelate. This analysis has been performed as part of this analysis of interrelation. Analysis of interrelation among such data will broaden the view on information and provide new insights, which allows a new approach to the utilization of retained information. **END.**

# V. Appendix

| Category | Appraisal Price (KRW) | Age | | | | | | | | | Total |
|---|---|---|---|---|---|---|---|---|---|---|---|
| | | 10s | 20s | 30s | 40s | 50s | 60s | 70s | 80s | 90s | |
| Analysis Targets | 0 million | 142 | 1,550 | 9,346 | 23,246 | 27,045 | 13,268 | 5,882 | 1,523 | 267 | 82,269 |
| | ~ 50 million | 1,573 | 4,492 | 14,982 | 36,682 | 36,343 | 18,084 | 9,373 | 2,150 | 728 | 124,407 |
| | 50 ~ 100 million | 616 | 3,355 | 14,758 | 31,032 | 24,841 | 10,741 | 4,779 | 999 | 302 | 91,423 |
| | 100 ~ 300 million | 579 | 3,802 | 19,273 | 46,374 | 40,392 | 15,131 | 6,285 | 1,250 | 371 | 133,457 |
| | 300 ~ 600 million | 153 | 892 | 5,178 | 16,430 | 17,494 | 6,786 | 2,417 | 511 | 104 | 49,965 |
| | 600 million ~ 1 billion | 39 | 276 | 1,446 | 4,935 | 6,206 | 2,665 | 997 | 198 | 22 | 16,784 |
| | 1~2 billion | 29 | 134 | 837 | 2,659 | 3,689 | 1,726 | 626 | 109 | 28 | 9,837 |
| | 2 billion ~ | 9 | 73 | 353 | 1,367 | 1,852 | 923 | 362 | 70 | 15 | 5,024 |
| Number of Deaths | 0 million | - | - | 37 | 163 | 280 | 239 | 309 | 140 | 24 | 1,192 |
| | ~ 50 million | 2 | 16 | 74 | 457 | 700 | 734 | 838 | 400 | 44 | 3,265 |
| | 50 ~ 100 million | - | 9 | 78 | 371 | 457 | 406 | 403 | 183 | 13 | 1,920 |
| | 100 ~ 300 million | 1 | 5 | 65 | 382 | 614 | 429 | 438 | 197 | 28 | 2,159 |
| | 300 ~ 600 million | - | 1 | 22 | 116 | 229 | 185 | 147 | 79 | 10 | 789 |
| | 600 million ~ 1 billion | - | 1 | 2 | 44 | 70 | 58 | 60 | 31 | 1 | 267 |
| | 1~2 billion | - | - | 2 | 22 | 40 | 44 | 41 | 16 | 5 | 170 |
| | 2 billion ~ | - | 1 | 1 | 7 | 23 | 15 | 15 | 9 | 1 | 72 |
| Potential Risk (Times) | 0 million | - | - | 1.8 | 1.4 | 1.0 | 0.8 | 0.8 | 0.4 | 0.2 | 0.7 |
| | ~ 50 million | 2.6 | 2.6 | 2.0 | 2.2 | 1.7 | 1.6 | 1.2 | 0.8 | 0.1 | 1.2 |
| | 50 ~ 100 million | - | 2.1 | 2.4 | 2.3 | 1.7 | 1.6 | 1.2 | 0.8 | 0.1 | 1.3 |
| | 100 ~ 300 million | 3.6 | 1.1 | 1.6 | 1.7 | 1.4 | 1.2 | 1.0 | 0.7 | 0.1 | 1.1 |
| | 300 ~ 600 million | - | 0.9 | 2.1 | 1.5 | 1.3 | 1.2 | 0.9 | 0.7 | 0.2 | 1.0 |
| | 600 million ~ 1 billion | - | 3.0 | 0.7 | 1.9 | 1.1 | 1.0 | 0.9 | 0.7 | 0.1 | 1.0 |
| | 1~2 billion | - | - | 1.1 | 1.8 | 1.1 | 1.2 | 1.0 | 0.6 | 0.3 | 1.0 |
| | 2 billion ~ | - | 11.4 | 1.3 | 1.1 | 1.2 | 0.7 | 0.6 | 0.5 | 0.1 | 0.7 |

Table 7: Ratio of Mortality Rate of Interest and Potential Risk of Targets of Interest
by Age and Appraisal Price